\documentclass[twocolumn,prb,showpacs,citeautoscript,floatfix]{revtex4}
\usepackage{amsfonts}
\usepackage{amssymb}
\usepackage{amsmath}
\usepackage{graphicx}
\usepackage{dcolumn}   
\usepackage{bm}        

\newcommand{\etal}{\mbox{\textit{et al.}}}

\begin{document}


\title{Equation of state and phonon frequency calculations of
diamond at high pressures.}

\author{K. Kunc}
\altaffiliation[Permanent address: ]{Laboratoire d'Optique des Solides,
CNRS and Universit\'{e} Pierre and Marie Currie, T13-C80, 4 pl. Jussieu,
F-75252 Paris - Cedex 05, France}
\author{I. Loa}
\author{K. Syassen}
\email[Corresponding author:~E-mail~]{k.syassen@fkf.mpg.de}
\affiliation{Max-Planck-Institut f{\"u}r Festk{\"o}rperforschung,
Heisenbergstrasse 1, D-70569 Stuttgart, Germany}

\date{29 April 2003}

\begin{abstract}
The pressure-volume relationship and the zone-center optical phonon
frequency of cubic diamond at pressures up to 600~GPa have been
calculated based on Density Functional Theory within the Local
Density Approximation and the Generalized Gradient Approximation.
Three different approaches, viz.\ a pseudopotential method applied
in the basis of plane waves, an all-electron method relying on
Augmented Plane Waves plus Local Orbitals, and an intermediate
approach implemented in the basis of Projector Augmented Waves have
been used. All these methods and approximations yield consistent
results for the pressure derivative of the bulk modulus and the
volume dependence of the mode Gr{\"u}neisen parameter of diamond. The
results are at variance with recent precise measurements up to
140~GPa. Possible implications for the experimental pressure
determination based on the ruby luminescence method are discussed.
\end{abstract}

\bigskip


\maketitle

\section{Introduction}

Diamond is the archetype of the covalently bonded, tetrahedrally
coordinated insulators. It's extreme hardness is highly valued in
technology and is also exploited in high pressure research when using
the diamond anvil cell. The elastic properties of diamond near ambient
conditions are well characterized through ultrasonic and Brillouin
techniques. However, as a consequence of the small compressibility, the
changes in elastic properties under high hydrostatic pressure are not
well confined experimentally. This, for instance, applies to the
variation of the bulk modulus $B$ with pressure $P$, $B_0^{\prime} =
(dB/dP)_{P=0}$, a basic parameter in the equation-of-state (EOS)
modelling. A property closely related to the compression behavior of
diamond is the pressure-induced frequency shift of the threefold
degenerate F$_{2g}$ zone-center optical phonon mode; its shift with
pressure provides an approximate measure of the change in relative
density, because the mode Gr{\"u}neisen parameter is close to
one\cite{Mitra69,Whalley76,Grims78,Hanfl85b}.

High-pressure x-ray diffraction
experiments\cite{Aleks87,Gille99,Occelli02} yield the ambient-pressure
bulk modulus $B_0$ in good agreement with acoustic
measurements\cite{McSki72,Grims75,Ramda93,Vogel96,Zoubo98}. The
analysis of diffraction data is usually based on adopting
$B_0^{\prime}=4.0$ obtained from ultrasonic measurements up to 0.2
GPa\cite{McSki72}. The only exception is the recent diffraction study
of diamond to 140~GPa by Occelli \etal\cite{Occelli02}, who report
$B_0^{\prime}=3.0$,  at variance with the ultrasonic measurement. It
was subsequently argued\cite{Holza02} that the ruby pressure
calibration\cite{Mao86} employed in the diffraction work of
Ref.~[\onlinecite{Occelli02}] may need a revision. If so, the pressure
shift of the F$_{2g}$ phonon frequency, also studied by Occelli
\etal~up to 140~GPa, should be affected in a similar manner. The
F$_{2g}$ phonon mode behavior at high pressures was frequently studied
by
Raman spectroscopy\cite{Mitra69,Whalley76,Grims78,Hanfl85b,Boppa85,%
Aleks87,Gonch85,Aleks86,Tardi90,Muino94,Schif97} and its possible role
in pressure calibration was addressed early
on\cite{Hanfl85b,Aleks87,Sherm85}.

We report the calculation of the EOS and optical phonon frequency of
diamond at high pressures within Density Functional Theory (DFT).
Extensive theoretical work on diamond under pressure by \textit{ab
initio} methods has addressed changes in bonding, elasticity, lattice
dynamics, thermodynamical properties, phase stability, and electronic
excitations, see, e.g., Refs.
[\onlinecite{Zunger77,Yin83,Cheli84,Cohen85,Cheli87,Hanfl85b,Niels86,Cardo86,Fahy87,Fahy87a,%
Fahy87b,Fahy88,Mailh91,Vanca92,Pavon93,Willa94,Clark95,Scand96,%
Serra98,Xie99,Wu99,Zhao99}]. Here, we are interested in a specific
question: What are, based on different implementations and
approximations of DFT, the constraints on $B_0^{\prime}$ of diamond
and on the nonlinear pressure shift of the optical phonon
frequency. The calculated values of $B_0^{\prime}$ reported in the
literature scatter by about 25\%, spanning a range similar to that
of the experimental results. Calculated pressure effects on the
optical phonon frequency were reported in Refs.
[\onlinecite{Hanfl85b}] and [\onlinecite{Niels86}]. Results of a
more recent calculation\cite{Wu99} show some disagreement with
Refs. [\onlinecite{Hanfl85b}] and [\onlinecite{Niels86}]. In view
of new experimental results\cite{Occelli02} and the -- at least
apparent -- uncertainties in the previous theoretical predictions
we considered it worthwhile to revisit the calculation of the EOS
and optical phonon frequency of diamond under pressure, combined
with accurate procedures to extract the parameters of interest.

\section{Details of the calculations}

The total energy calculations performed in this work are based on
DFT\cite{Hohenberg64} within, on one hand, the Local Density
Approximation\cite{Kohn65} (LDA) and, on the other hand, the
Generalized Gradient Approximation\cite{PW92,PBE96} (GGA). In order to
grasp the uncertainties consequent to the choice of the computational
method and of its inherent assumptions, we are using simultaneously
three different approaches, viz. the pseudopotential method applied in
the basis of plane waves\cite{Gonze02,Goe97,PTAA92,Gon96} (PW), an
all-electron method relying on the Augmented Plane Waves plus Local
Orbitals\cite{soft:WIEN2k,SNS00,MBSS01} (APW+lo), and an intermediate
approach implemented in the basis of Projector Augmented
Waves\cite{soft:VASP-short,Kress99} (PAW); the latter
approach\cite{Bloch94} treats the valence states as part of an
all-electron problem and describes them by all-electron wave-functions.

The pseudopotential employed is the ``dual-space'' separable
pseudopotential of Hartwigsen, Goedecker, and Hutter\cite{Hartw98}
and its GGA counterpart constructed by X. Gonze\cite{Gonze02,Gonze};
the PAW potentials were constructed by G. Kresse and denoted as C\_h
in Ref. [\onlinecite{soft:VASP-short}]. All methods take into
account scalar-relativistic corrections (though no meaningful
contribution expected in diamond), either explicitly (APW+lo) or
through the construction of the (pseudo-)potentials.

The numerical convergence of all three methods, with respect to the
size of the basis set and k-space sampling was thoroughly tested.
The plane-wave cutoffs of 75 Hartree and 1500 eV were applied with,
respectively, the PW and the PAW-basis. These rather large values
are required for getting reliable results for the \textit{pressure}
which is calculated analytically, in both approaches, with the aid
of the stress theorem\cite{Nielsen83,Nielsen85}. In the APW+lo
method a muffin-tin radius of $R_{MT} = 1.2$~a.u.\ was used. The
plane-wave expansion was limited by $R_{MT} \times K_{max} = 9$ and
the charge density Fourier expansion by $G_{max} = 20$~Ry$^{1/2}$.
In the APW+lo method the $P(V)$ results were generated from
$E^{\text{tot}}(V)$ fitted as described below.

The integration over the Brillouin zone is performed using the same set
of 28 k-points in the irreducible wedge in both plane-wave based
approaches; the set was generated by the ``special points'' approach
\cite{Monkh76} using a $6 \times 6 \times 6$ mesh with 4 different
fractional shifts. In the APW+lo approach the tetrahedron integration
was employed\cite{JA71,BJA94}, based on  a uniform, $\Gamma$-centered
$10 \times 10 \times 10$ mesh of 47 k-points.

The frozen-phonon approach, which consists in the evaluation of the
total energy $E^{\rm tot}$ of the crystal with frozen-in atomic
displacements, was applied (in the two plane-wave based methods only)
with the same cutoffs and with the same k-point mesh ($6 \times 6
\times 6$ + 4 shifts) which, due to the lowered symmetry, results in a
91 k-points set. Small displacements $\vec u(1) = \pm (u,u,u)$ and
$\vec u(2) = \mp (u,u,u)$ with $u/a$ = 0.002 were applied to the two
atoms of the basis -- thus either compressing or stretching the C(1) --
C(2) bond, like in the zone-center F$_{2g}$ mode -- and the two values
obtained for $\Delta E^{\rm tot}$ [viz. the $\Delta E^{\rm
tot}$(outward) and $\Delta E^{\rm tot}$(inward)] were averaged. The
eigenfrequency is then found from the expression for the energy of a
harmonic oscillator
\begin{equation}
\Delta E^{\rm tot} = \frac{1}{2}M \omega^2 \left[\vert \vec u(1)
\vert^2 + \vert \vec u(2) \vert^2\right]~,
\end{equation}
where $M$ is the atomic mass. The above-mentioned averaging procedure
of the $\Delta E^{\rm tot}$(outward) and $\Delta E^{\rm tot}$(inward)
eliminates the cubic contribution to $\Delta E^{\rm tot}$. It turns out
that the remaining quartic anharmonicity is small at $u/a = 0.002$ and
contributes to the uncertainty of the resulting eigenfrequencies by
less than 0.1 cm$^{-1}$ (checked by repeating the same calculations
with displacements $u/a$=0.0015 and 0.003).

Total energies, pressures, and phonon frequencies were calculated at 12
different volumes ranging from 6.35~\AA$^3$/atom down to
3.35~\AA$^3$/atom, i.e., for about 10\% volume expansion to 40\%
compression relative to the experimental equilibrium volume of
5.6725~\AA$^3$/atom at 300~K \cite{Sato02}. The lower volume limit
corresponds to roughly 600~GPa maximum pressure.

\begin{figure}[tb]\centering
\includegraphics[width=80mm,clip]{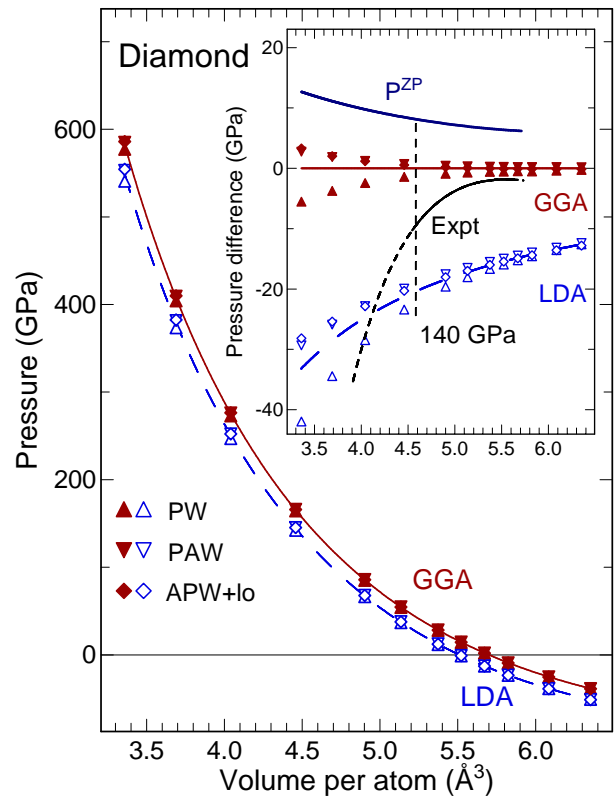}
\caption{\label{fig1} Pressure-volume results for diamond from
plane wave (PW, triangles up), projector augmented wave (PAW,
triangles down), and all-electron (APW+lo, diamonds) calculations.
Solid and dashed lines represent fits of Eq.~\ref{epv1} to the
combined GGA (filled symbols) and LDA (open symbols) results,
respectively.  The inset offers a zoomed view of pressure
differences relative to the average $P(V)$ relation for GGA
calculations, including the differences for experimental data of
Ref.~[\onlinecite{Occelli02}] and their extrapolation. The volume
dependence of the zero-point pressure $P^{ZP}(V)$ is also shown in
the inset.}
\end{figure}

\begin{figure}[tb]\centering
\includegraphics[width=80mm,clip]{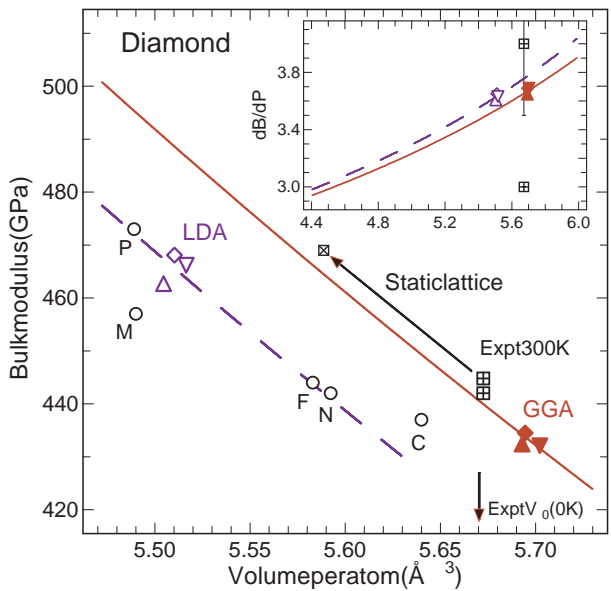}
\caption{\label{fig2}  Bulk modulus $B_0$ of diamond obtained from
different calculations plotted against the calculated equilibrium
volume $V_0$. Symbols for the present GGA and LDA results as in
Fig.~\ref{fig1}. Other calculated values of $B_0$
[\onlinecite{Cheli84,Niels86,Fahy87,Mailh91,Pavon93}] are indicated
by open circles (the attached letters refer to the first-author
name). The experimental values of $B_0$ originate from ultrasonic
measurements\cite{McSki72} and Brillouin scattering\cite{Vogel96}.
The solid (dashed) line refers to the volume dependence of the bulk
modulus for combined GGA (LDA) results. The inset shows the
pressure derivative of the bulk modulus versus atomic volume. The
calculated $B_0^{\prime}$ values and the two experimental ones are
represented by symbols.}
\end{figure}

\section{Pressure-volume relationship}

\subsection{Analytical form}

Figure~\ref{fig1} shows calculated pressures as a function of
volume. An excellent consistency of the data points obtained by the
three different methods is immediately apparent. Also, the present
results substantiate the claims made in Ref. [\onlinecite{Kress99}]
concerning the equivalence of the PAW approach with all-electron
methods such as that of Ref.~[\onlinecite{soft:WIEN2k}].

The proper analysis of the $P(V)$ results yields the equilibrium
properties $V_0$, $B_0 = - V_0 (dP/dV)_0$, and $B_0^{\prime} =
(dB/dP)_0 = (d \ln B/d \ln V)_0$ (throughout this manuscript the
subscript 'zero' refers to zero (ambient) pressure or equilibrium
volume, either calculated or experimental depending on context). One
also needs to identify an analytical form of the $P(V)$ behavior
which, using the above parameters, fits the calculated pressures
over the full volume range.

We have tried several of the common equation-of-state
forms\cite{Murna44,Birch78,Vinet86,Holza96}. The best analytical
expression, in the least-squares sense, was identified as
\begin{equation}\label{epv1}
P(V)  = 3 B_0 X^{-n}(1-X) \exp[\eta\,(1-X)]
\end{equation}
where $X = (V/V_{0})^{1/3}$, $\eta=3 B_0^{\prime}/2 + 1/2 - n$, and
$n=7/2$. With $n=7/2$ Eq.~\ref{epv1} is a blend of the
Vinet\cite{Vinet86} (or Rydberg\cite{Stace81,Gaura02}) form and the
Holzapfel\cite{Holza96} expression, for which $n=2$ and $n=5$,
respectively. Actually, the parameter $n$ varies between 3.4 and
3.6 for individual sets of calculated $P(V)$ data of diamond.
Without loss of significant digits in the fitted parameters of
interest, we fix $n$ at a value of 7/2; in this case the energy
versus volume relation, obtained by integration of Eq.~\ref{epv1},
can be written without invoking special functions other than the
error function:
\begin{eqnarray}\label{eev1}
E(V) &=& E_0 + 9\,B_0\,V_0 \left[f(V)-f(V_0)\right] \exp(\eta)/\sqrt{\eta} \\
f(V)&=&\sqrt{\pi}(2\eta+1)\,\mathrm{erf}(\sqrt{\eta}X^2)+2\sqrt{\eta}\exp(-\eta
X)/X^2~. \nonumber
\end{eqnarray}
Here, $E_0$ is the energy at $V=V_0$. Equation~\ref{epv1} was used
to fit the $P(V)$ results obtained via the stress theorem (PW and
PAW methods), while Eq.~\ref{eev1} served to determine the EOS
parameters from total energies (APW+lo method). For the PW and PAW
methods it was carefully checked that the directly calculated
$P(V)$ data are consistent with those obtained by differentiation
of $E(V)$. For all sets of calculated $P(V)$ or $E(V)$ data the rms
deviations were less than 0.04~GPa or 0.05~meV, respectively.

Aleksandrov \etal\cite{Aleks87} pointed out that a quadratic dependence
of pressure on change in relative density $\Delta \rho/\rho_0 =
V_0/V-1$, written as
\begin{equation} \label{epv2}
P(\rho) = B_0 \frac{\Delta \rho}{\rho_0} \left[ 1 +
\frac{B_0^{\prime}-1}{2} \frac{\Delta \rho}{\rho_0}\right]~,
\end{equation}
applies in the case of diamond for $P \leq 100$~GPa. This
expression is found to fit the calculated results up to 600~GPa
quite well. It is only in the statistical sense that Eq.~\ref{epv2}
is slightly inferior to Eq.~\ref{epv1}, i.e., the rms deviations
are larger by a factor of two, but this is not relevant for the
present discussion. The bulk modulus, obtained by differentiation
of Eq.~\ref{epv2} with respect to normalized volume, is given by
\begin{equation} \label{ebm2}
B(\rho) = B_0 \frac{\rho}{\rho_0} \left[1+(B_0^{\prime}-1)\frac{\Delta
\rho}{\rho_0}\right]~.
\end{equation}
\'{E}quation \ref{ebm2} corresponds to a quadratic dependence of the
bulk modulus on relative change in density.

\begin{table*}[tb]\centering
\caption{Calculated equation-of-state [Eq.~\ref{epv1}] and optical
phonon parameters of diamond. Selected experimental results are
listed in the lower part of the Table. The static-lattice entries
correspond to switching off (at $T$ = 0~K) the effect of zero point
motion. The calculated phonon pressure coefficients are obtained
via $(d\omega/dP)_0 =\gamma_0 \omega_0/B_0$ with parameters given
in the same row of the Table. Likewise, the experimental mode
Gr{\"u}neisen parameters were obtained from the experimental pressure
coefficients unless noted otherwise. In the row marked GGA\&LDA
both the ambient pressure volume and bulk modulus are normalized to
one.} \label{tab1}

\medskip

\begin{ruledtabular}
\begin{tabular}{lcccccccc}
Method & $V_0$    & $B_0$     &$B_0^{\prime}$&n& $\omega_0$&$\gamma_0$&$\gamma_0^{\prime}$& $(d\omega/dP)_0$\\
       & \AA$^3$   & GPa       &&            &cm$^{-1}$& &     &cm$^{-1}$/GPa\\
\hline
All LDA &5.510(5)  &465(3)  &3.63(3) &7/2&1322(2)&1.003(3)&0.79(5)&2.87\\
All GGA &5.697(4)  &433(2)  &3.67(3) &7/2&1290(2)&0.995(3)&0.80(5)&3.00\\
GGA\&LDA normalized &1  &1  &3.65(5) &7/2&1&1.000(5)&0.80(5)&\\
\hline
Expt. $T$=300~K & 5.6725\footnotemark[1] & 442\footnotemark[2] &
4.0(5)\footnotemark[2] &&1332.5&0.962(15)\footnotemark[3] &&2.90(5)\footnotemark[4]\\
&  & 444.8(8)\footnotemark[5] && &1332.40(5)\footnotemark[5] &
 1.00(3)\footnotemark[6]&&3.00(10)\footnotemark[6]\\
& 5.674(1)\footnotemark[7] & 446(1)\footnotemark[7]
& 3.0(1)\footnotemark[7] &2&1333\footnotemark[7]&$\overline{\gamma}$=0.97\footnotemark[7] &&2.83\footnotemark[7]\\
\hline
Expt. $T\rightarrow$0~K & 5.6707\footnotemark[1] &445\footnotemark[9]&& &1332.70(3)&& &\\
Static lattice & 5.5886\footnotemark[8] & 469\footnotemark[9]&  & &&& &\\
          & 5.6122\footnotemark[10] & 462\footnotemark[9]  && &&&&\\
\end{tabular}
\end{ruledtabular}

\footnotetext[1]{Ref.~[\onlinecite{Sato02}].}

\footnotetext[2]{Ref.~[\onlinecite{McSki72}]; ultrasonic experiments up
to 0.2~GPa.}

\footnotetext[3]{Experimental mode Gr{\"u}neisen parameters reported in the
literature
[\onlinecite{Mitra69,Whalley76,Grims78,Hanfl85b,Boppa85,Aleks86,Aleks87,Tardi90}]
vary between 0.90 to 1.06.}

\footnotetext[4]{'Best' value in the literature according to
Ref.~[\onlinecite{Schif97}].}

\footnotetext[5]{Ref.~[\onlinecite{Vogel96}]; Brillouin and Raman
scattering.}

\footnotetext[6]{Value obtained for a revised linear pressure
coefficient of the ruby line shift as explained in the text.}

\footnotetext[7]{Ref.~[\onlinecite{Occelli02}]; x-ray diffraction
(Vinet fit) and Raman scattering up to 140~GPa. The mode Gr{\"u}neisen
parameter $\overline{\gamma}$ is the average for the pressure range
0--140 GPa.}

\footnotetext[8]{Ref.~[\onlinecite{Herre00}]; the classical limit in
path integral Monte Carlo simulations.}

\footnotetext[9]{Estimated value, see text.}

\footnotetext[10]{Ref.~[\onlinecite{Vogel96}]; based on extrapolation
of isotope effects.}

\end{table*}

\subsection{Calculated EOS parameters}

The obtained pairs of ($V_0$,$B_0$) values are displayed in
Fig.~\ref{fig2}. Within a given approximation for the
exchange-correlation functional (GGA or LDA), the three methods
employed here (PW, PAW, APW+lo) yield results for $V_0$ and $B_0$
which are in very good agreement with each other. The choice of the
exchange-correlation functional, however, leads to differences of
4\% in $V_0$ and 5\% in $B_0$, reflecting in part the well-known
overbinding of the LDA. All $B_0^{\prime}$ values (cf. inset to
Fig.~\ref{fig2}) fall into a narrow range between 3.6 and 3.7 , the
difference between averaged LDA and GGA results being close to 1\%.

Since the calculated EOS parameters clearly split into two groups,
i.e. the GGA and LDA results, with nearly identical parameter
values within each group, we have combined the $P(V)$ points for
GGA and LDA, respectively, to obtain the GGA and LDA parameters
$V_0$, $B_0$, and $B_0^{\prime}$ listed in Table~\ref{tab1}. The
corresponding $P(V)$ relations are shown by solid and broken lines
in Fig.~\ref{fig1}, where the inset illustrates deviations of
individual calculated results from the average curves. Similarly,
the solid and broken lines in Fig.~\ref{fig2} and its inset show
the corresponding volume dependences of the bulk modulus and its
pressure derivative.

The larger $B_0$ value for LDA compared to GGA correlates with the
smaller equilibrium volume. However, the $B(V)$ curve for LDA falls
below that for GGA (see Fig.~\ref{fig2}). This also applies to $B_0$
values obtained in other self-consistent LDA calculations for diamond
\cite{Niels86,Fahy87,Mailh91,Pavon93} (represented by open circles in
Fig.~\ref{fig2}).

\subsection{Zero-point motion effects}

The calculated results for $V_0$ and $B_0$ should be compared to
experimental properties after 'correction' for vibrational effects.
Thermal effects are almost negligible at 300~K because of the high
Debye temperature of diamond ($\sim$2000 K), but zero-point motion
needs to be considered. According to Monte Carlo
simulations\cite{Herre00} the static-lattice equilibrium volume for
diamond is about 5.59~\AA$^3$. All the data for $V_0$ shown in
Fig.~\ref{fig2} fall into a range of $\pm 2$\% around this value. The
static-lattice bulk modulus at $V=5.59$~\AA$^3$ is estimated to be
469~GPa (this follows from the experimental values of $V_0$ and $B_0$
and the calculated $B_0^{\prime}$). Thus, relative to static-lattice
properties, the GGA results cannot be considered superior to the LDA
ones; they appear better when compared to experimental data because the
GGA errors happen to mimic the zero-point effects.

The zero-point vibrational pressure $P^{ZP}$, i.e., the isochoric
change in pressure when zero-point motion is switched on, is about
6~GPa at the static-lattice value of $V_0$\cite{Herre00}. An
approximate relation for the volume dependence of $P^{ZP}$ is
\begin{equation*}
P^{ZP}(V) \approx - \frac{dP(V)}{dV} \Delta V^{ZP}(V)~,
\end{equation*}
where $\Delta V^{ZP}$ is the isobaric volume expansion due to
zero-point motion. For getting $\Delta V^{ZP}(V)$ within the
quasi-harmonic thermodynamics, we simply refer to the calculated
pressure dependence of the zero-point expansion presented in Fig.~10 of
Ref.~[\onlinecite{Herre00}]. The corresponding volume dependence of
$P^{ZP}$ is displayed in the inset of Fig.~\ref{fig1}. $P^{ZP}$
increases by about 6~GPa at 600~GPa. Adding such small \textit{changes}
in $P^{ZP}$ to the calculated $P(V)$ results leads to an increase of
the $B_0$ values by about 1\%, but would not affect the calculated
$B_0^{\prime}$ values (within the estimated uncertainty of the optimum
value given below).

\subsection{The essence of the EOS calculations}

\begin{figure}[tb]\centering
\includegraphics[width=80mm,clip]{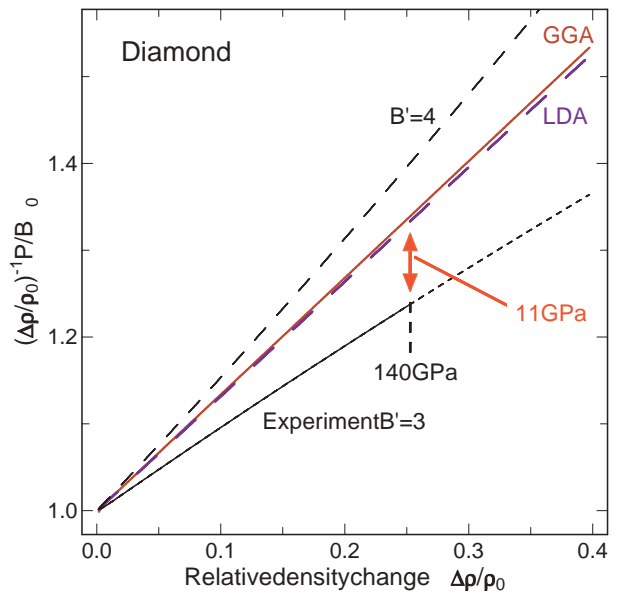}
\caption{\label{fig3} Calculated pressure-volume relations of
diamond in reduced coordinates and using $V_0$, $B_0$, and
$B_0^{\prime}$ from Table~\ref{tab1}. Experimental data of Ref.
[\onlinecite{Occelli02}] are shown for comparison. These
experiments cover a pressure range of about 140~GPa. The line
labelled $B_0^{\prime}=4$ represents the EOS with parameters $B_0$
and $B_0^{\prime}$ from ultrasonic experiments
[\onlinecite{McSki72}].}
\end{figure}

In Fig.~\ref{fig3}, the calculated $P(V)$ relations are plotted in
reduced coordinates, i.e. pressure normalized by $B_0\,(\Delta
\rho/\rho_0)$ as a function of $\Delta \rho/\rho_0$. In this
representation small differences in ambient-pressure volumes and
bulk moduli are suppressed and a linear slope corresponds to a
quadratic dependence of pressure on change in relative density
(Eq.~\ref{epv2}). The LDA and GGA results hardly differ in slope.
In this way, Fig.~\ref{fig3} illustrates the main information we
extract from our EOS calculations for diamond:

(1) The value of $B_0^{\prime}$ does not depend much on crystal
potential and basis set issues and on approximations for the
exchange-correlation functional. Our average value is $$B_0^{\prime} =
3.65 \pm 0.05,$$ where the small uncertainty reflects the scatter for
the different methods of calculation.

(2) Within the volume range considered here, the most appropriate
3-parameter analytical form for the $P(V)$ relation of diamond
(Eq.~\ref{epv1} or \ref{epv2}) is transferable between LDA and GGA
solutions; only $V_0$ and $B_0$ need to be adjusted.

(3) Inserting the calculated $B_0^{\prime}$ value and the
experimental data for $V_0$ and $B_0$ (see Table~\ref{tab1}) into
Eq.~\ref{epv1} (Eq.~\ref{epv2}) is considered to yield, on the
basis of this work, the optimum representation of the EOS of
diamond at 300~K. Actually, at the experimental $V_0$ the GGA bulk
modulus (calculated value of 440~GPa plus 1\% correction for
zero-point motion effects) happens to be very close to the
experimental $B_0$, and the calculated pressure at the experimental
$V_0$ is only 2~GPa. Therefore, the optimum EOS corresponds to the
$P(V)$ relation obtained from GGA calculations, with only a small
correction applied.

\subsection{Comparison with other results for $B_0^{\prime}$}

Calculated values of $B_0^{\prime}$ reported in the literature are
3.54 [\onlinecite{Cheli84}], 3.6, 4.5 [\onlinecite{Niels86}], 3.24
[\onlinecite{Fahy87}], 3.5 [\onlinecite{Pavon93}] and 3.97
[\onlinecite{Herre00}]. In some cases, the differences to our
result are small. Larger deviations may in part result from the
procedure used to extract the EOS parameters (e.g., fit of a
Murnaghan equation which is inadequate).

The experimental value $B_0^{\prime}$ =4 $\pm$ 0.5
(Ref.~[\onlinecite{McSki72}]) derived from sound speed measurements at
low pressures (0.2~GPa) is not sufficiently accurate to test the
calculated result.

The only other experimental value of $B_0^{\prime}$ stems from the
recent x-ray diffraction experiments up to 140~GPa\cite{Occelli02}.
The EOS data appear to be of high quality. They were measured using
helium as a pressure medium which is considered to provide almost
hydrostatic conditions in DAC experiments. The obtained value of
$B_0^{\prime} = 3.0(1)$ is inconsistent with the ultrasonic result
and also significantly smaller than the calculated value. The
latter difference is illustrated in Fig.~\ref{fig3}. At nominally
140~GPa, about the maximum pressure reached in the experiments, the
difference between our calculated $B_0^{\prime}$ value and that
Ref.~[\onlinecite{Occelli02}] translates to a pressure difference
of about +11 GPa. This means that either repulsion is slightly
overestimated in the calculations (independent of the
exchange-correlation functional) or that the experimental data
suffer from systematic errors, a combination of these effects not
being ruled out. We will return to this issue below.

\section{Optical phonon frequency}

\subsection{Calculated phonon frequency versus volume}

\begin{figure}[t!]\centering
{\includegraphics[width=80mm,clip]{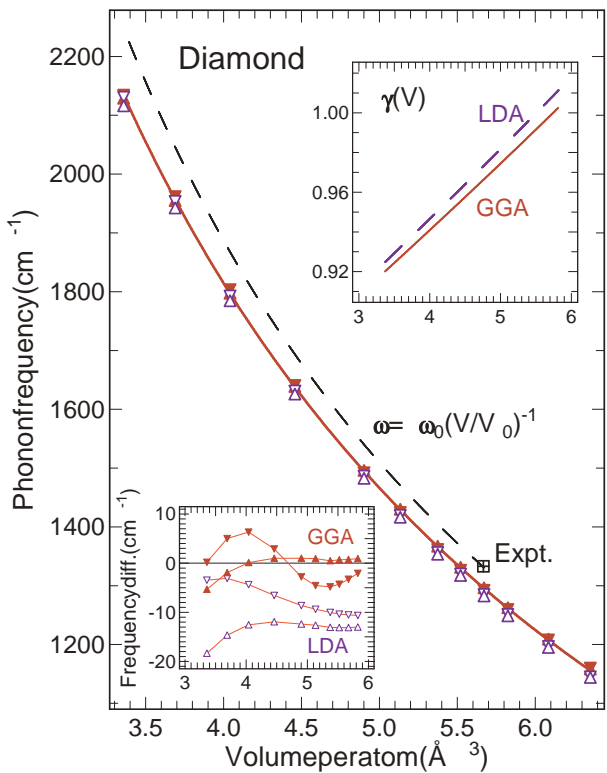}} \caption{\label{fig4}
Calculated zone-center optical phonon frequency of diamond as a
function of atomic volume. The solid line guides through the
combined GGA results. A Gr{\"u}neisen relation Eq.~\ref{epgr} with
parameters $\omega_0 = 1333$~cm$^{-1}$ (experimental phonon
frequency) and $\gamma_0=1$ is indicated for comparison (dashed
line). Differences of individual calculated results with respect to
the 'combined GGA' curve are shown in an inset. The second inset
illustrates the calculated volume dependence of the mode Gr{\"u}neisen
parameter (for combined GGA and LDA results, respectively).}
\end{figure}

Figure~\ref{fig4} shows calculated F$_{2g}$ phonon frequencies as a
function of volume. At a given volume, the frequencies of all
calculations are quite consistent with each other. The frequencies are
too low by about 3\% if compared to the experimental phonon frequency
at ambient pressure. This small discrepancy does not originate from
anharmonicity connected to the frozen-phonon displacement.

The volume dependence of phonon frequencies is usually characterized
by the simple scaling law
\begin{equation}\label{epgr}
\frac{\omega(V)}{\omega_0} = \left(\frac{V}{V_0}\right)^{-\gamma}=
\left(\frac{\rho}{\rho_0}\right)^{\gamma} ~,
\end{equation}
which assumes that the mode Gr{\"u}neisen parameter $\gamma = - d \ln
\omega/d \ln V$ is independent of volume. The frozen-phonon results
indicate some volume dependence of $\gamma$, an observation which was
also noted by Nielsen\cite{Niels86}. The expression
\begin{equation}\label{enuv}
\frac{\Delta
\omega(V)}{\omega_0} = \frac{\gamma_0}{\gamma_0^{\prime}}
\left[\left(\frac{V}{V_0}\right)^{-\gamma_0^{\prime}}-1\right]~.
\end{equation}
with $\Delta\omega(V)=\omega(V)-\omega_0$ yields a slightly better
match of our frozen-phonon results (3~cm$^{-1}$ rms deviation or better
for individual sets). Within standard deviations, the parameters
$\gamma_0$ and $\gamma_0^{\prime}$ for the combined GGA and LDA results
are identical (see Table~\ref{tab1}). The upper inset to
Fig.~\ref{fig4} illustrates the small volume dependence of the mode
Gr{\"u}neisen parameter resulting from $\gamma_0^{\prime} < 1$.

With $\rho/\rho_0 \approx \omega/\omega_0$ ($\gamma\approx 1$) it
follows from Eq.~\ref{ebm2} that the ratio of normalized bulk
modulus to normalized phonon frequency is approximately linear in
relative density and the slope is $B_0^{\prime}-1$:
\begin{equation}
\frac{B}{B_0}  \left(\frac{\omega}{\omega_0}\right)^{-1} \approx 1 +
(B_0^{\prime}-1)\frac{\Delta \rho}{\rho_0}~. \end{equation}
This scaling between bulk modulus and optical phonon frequency
could come handy as a simple relation between an elastic and a
dynamical property of diamond.

\subsection{Calculated vs. experimental mode Gr{\"u}neisen parameter}

To compare the calculated phonon results with experimental data, we
first consider the measured linear pressure coefficient of the
phonon frequency. According to Schiferl \etal\cite{Schif97}, the
'best' experimental value in the literature, taken as the average
of Refs. [\onlinecite{Hanfl85b,Boppa85,Muino94}], should be
$(d\omega/dP)_0=2.90\pm0.05$~cm$^{-1}$/GPa ($\gamma_0\approx0.96$).
This value originates from experiments which employed the ruby
pressure calibration with a linear coefficient $A = (dP/d \ln
\lambda)_0 = 1905(10)$~GPa for the R1 line wavelength ($\lambda$)
shift near zero pressure as determined by Piermarini
\etal\cite{Pierm75}~  Recent high-precision measurements of the
ruby line shift up to 1~GPa\cite{Grass01} and a
reinterpretation\cite{Holza02} of R1 line shift data measured up to
$\sim$20~GPa\cite{Pierm75,Nakan00} indicate that the parameter $A$
is smaller ($A = 1820 \pm 30$~GPa) compared to the previously
accepted value. This revision leads to a corresponding increase of
the phonon pressure coefficient. In this context it is helpful that
the ratio of the diamond phonon frequency shift to ruby wavelength
shift is explicitly given in Refs. [\onlinecite{Hanfl85b}] and
[\onlinecite{Aleks87}]. The results, 7.94(10)~cm$^{-1}$/nm and
7.85(6)~cm$^{-1}$/nm, agree quite well with each other. Taking the
average value (7.9(1)~cm$^{-1}$/nm) in combination with $A$ =
1820(30)~GPa, the corrected value from DAC experiments should be
$(d\omega/dP)_0=3.0(1)$~cm$^{-1}$/GPa. Within experimental
uncertainties, this linear pressure coefficient agrees with results
obtained by methods which do not involve any ruby
calibration\cite{Mitra69,Whalley76,Grims78}, in particular that of
Whalley \etal\cite{Whalley76}~ With $B_0$=444.8 GPa the
corresponding mode Gr{\"u}neisen parameter becomes $\gamma_0=1.00(3)$,
the error being mainly due to the uncertainty of the ruby
coefficient $A$. We note the excellent agreement of the corrected
experimental value for $\gamma_0$ with the results obtained within
the LDA and GGA approximations (see Table~\ref{tab1}).

Occelli \etal~report an \textit{average} mode Gr{\"u}neisen parameter of
$\overline{\gamma}=0.97$ for the range 0--140 GPa. It should be pointed
out that their value of $\overline{\gamma}$ is independent of the
pressure scale used in the experiments because the volume was measured
directly. Taking into account the predicted volume dependence
$\gamma(V)$ (see inset to Fig.~\ref{fig4}), the calculations are fully
consistent with the experimental volume dependence of the phonon
frequency. This is no longer the case if we turn to the pressure
dependence.

\subsection{The pressure -- phonon-frequency relationship}

One can solve Eq.~\ref{enuv} for the relative volume, insert the
corresponding expression into Eq.~\ref{epv1} (Eq.~\ref{epv2}), and
use the 'GGA' and 'LDA' parameters from Table~\ref{tab1} to obtain
an analytical form for pressure as a function of phonon frequency.
On the other hand, in the analysis of earlier phonon
calculations\cite{Hanfl85b,Niels86}, the $P(\omega)$ behavior was
parametrized using the analog of a Birch expression\cite{Birch78}
\begin{equation}\label{ebirch}
P(X) = \frac{3}{2} b_0 (X^7 - X^5) [1 + \eta (1-X^2)]
\end{equation}
with $X=(\omega/\omega_0)^{1/3}$. Equation~\ref{ebirch} (more or
less an \textit{ad hoc} choice in the earlier work) happens to
yield excellent representations of the present calculated
$P(\omega)$ results. To facilitate direct comparison with
Refs.~[\onlinecite{Hanfl85b}] and [\onlinecite{Niels86}], the
parameters $b_0$ and $\eta$ are summarized in Table~\ref{tab2}.
Note that for our calculated data $\eta$ is the only adjustable
parameter, because $b_0 = B_0/\gamma_0$.

\begin{table}[tb!]\centering
\caption{Calculated zone-center optical phonon properties of
diamond. The parameters for the pressure versus frequency relation,
Eq.~\ref{ebirch}, are given. Related experimental data are listed
in the last two rows of the Table. Note that the linear pressure
coefficient and the parameter $b_0$ are related by $(d\omega/d P)_0
= \omega_0/b_0 = \gamma_0 \omega_0/B_0$.} \label{tab2}
\medskip
\begin{ruledtabular}
\begin{tabular}{lcccc}
Method or     & $\omega_0$&$b_0$&$\eta$&$(d\omega/dP)_0$\\
Source         &cm$^{-1}$& GPa &&cm$^{-1}$/GPa\\
\hline
All GGA& 1290&429(6)&0.045(50)&3.00\\
All LDA&  1322&460(3)&0.090(20)&2.87\\
Hanfland \etal\footnotemark[1] &1341&480&0.118&2.79\\
Nielsen\footnotemark[2] &1306&456&0.066&2.86\\
\hline
Experiment &1333&460& - &2.90(5)\footnotemark[3]\\
Experiment &1333&446& - &3.00(10)\footnotemark[4]\\
\end{tabular}
\end{ruledtabular}
\footnotetext[1]{Ref.~[\onlinecite{Hanfl85b}].}

\footnotetext[2]{Ref.~[\onlinecite{Niels86}].}

\footnotetext[3]{Refs. [\onlinecite{Schif97}].}

\footnotetext[4]{Corrected value based on revised linear pressure
coefficient of the ruby R1 line shift.}

\end{table}

\begin{figure}[tb]\centering
\includegraphics[width=80mm,clip]{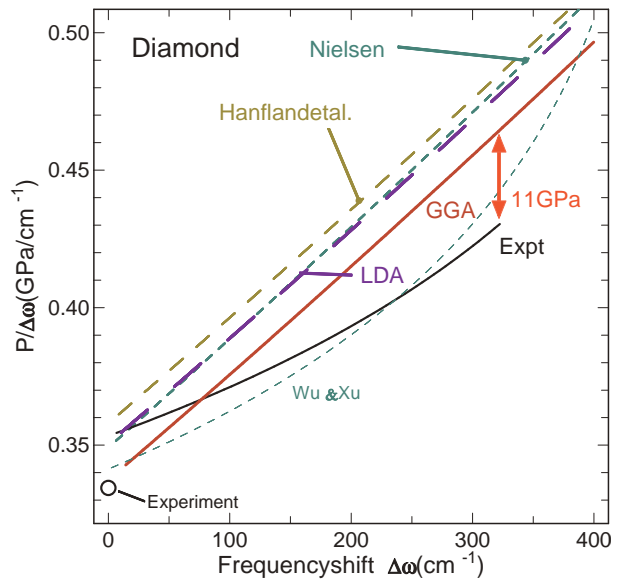}
\caption{\label{fig5} Calculated pressure as a function of optical
phonon frequency shift for diamond. A pressure range of about
200~GPa is covered. Pressure is divided by the frequency shift in
order to illustrate the nearly quadratic dependence on frequency
shift. In this representation the intercept at zero frequency shift
corresponds to the inverse linear pressure coefficient $(d
\omega/dP)^{-1}$. The open circle refers to the corrected
experimental value. Besides the present LDA and GGA results, the
figure shows the calculated results of Hanfland
\etal\cite{Hanfl85b}, Nielsen\cite{Niels86}, and
Wu~\&~Xu\cite{Wu99}. The experimental data of Occelli
\etal\cite{Occelli02} are shown for comparison.}
\end{figure}

Applicability of Eq.~\ref{epv2} in combination with $\gamma \approx
1$ implies a nearly quadratic dependence of pressure on change in
phonon frequency. Thus, we compare calculated and experimental
results in terms of reduced coordinates $P/\Delta \omega$ and
$\Delta \omega$, cf. Fig.~\ref{fig5}. Most of the calculated
results represented in Fig.~\ref{fig5}, i.e., the present ones
within LDA and GGA and those of Refs.~[\onlinecite{Hanfl85b}] and
[\onlinecite{Niels86}], agree very well with respect to the slope.
This is of course reflected in the small scatter of the $\eta$
values given in Table~\ref{tab2}. The average value from our
calculations is
$$\eta = 0.068 \pm 0.03~.$$
Inserting this value and the experimental data for $\omega_0$ and
$b_0 = \omega_0 (dP/d\omega)_0$ (Table~\ref{tab2}) into
Eq.~\ref{ebirch} yields the $P(\omega)$ dependence consistent with
experimental data near ambient pressure and with the nonlinear
behavior predicted by the present calculations.

In the reduced coordinates of Fig.~\ref{fig5} the calculated
$P(\omega)$ results of Wu and Xu\cite{Wu99} do not agree well with
those of other calculations. They use the quadratic function
\begin{equation}\label{eqwp}
\omega(P) = \omega_0 + a_1\,P + a_2\, P^2
\end{equation}
to fit their data. The choice of pressure as the independent
variable in a quadratic expression is not appropriate in view of
Eq.~\ref{epv2} and $\gamma \approx 1$. It leads to parameter
correlation which is the possible reason that their results exhibit
curvature in Fig.~\ref{fig5}.

Occelli \etal\cite{Occelli02} have measured the optical phonon
frequency for pressures up to 140~GPa. They also give frequency as a
quadratic function of pressure, i.e, $a_1 = 2.83$~cm$^{-1}$/GPa and
$a_2=-3.65 \times 10^{-3}$~cm$^{-1}$/GPa$^2$ in Eq.~\ref{eqwp}. The
curvature of their data in Fig.~\ref{fig5} could again be related to
the particular choice of the fitting function. The main observation,
however, is that the average slope for the experimental data differs
from the one predicted by theory. At the frequency corresponding to
140~GPa experimental pressure the pressure difference between
calculations and experiment again amounts to about 11~GPa.

The sign and magnitude of this difference is very similar to that
encountered when comparing experimental and calculated EOS results.
Thus, in terms of pressure we find differences between calculated and
experimental results which are about the same for different physical
quantities considered. This hints to an explanation where the
discrepancies are caused by the same systematic error(s).

\section{Remarks on the ruby calibration}

We consider the possibility that the discrepancy in the high pressure
regime between experimental and calculated EOS and phonon frequency
results for diamond are caused by some error in the experimental
pressure scale, i.e. the calibration of the ruby R1 line shift
according to Ref. [\onlinecite{Mao86}].

\begin{figure}[tb]
\centerline{\includegraphics[width=80mm,clip]{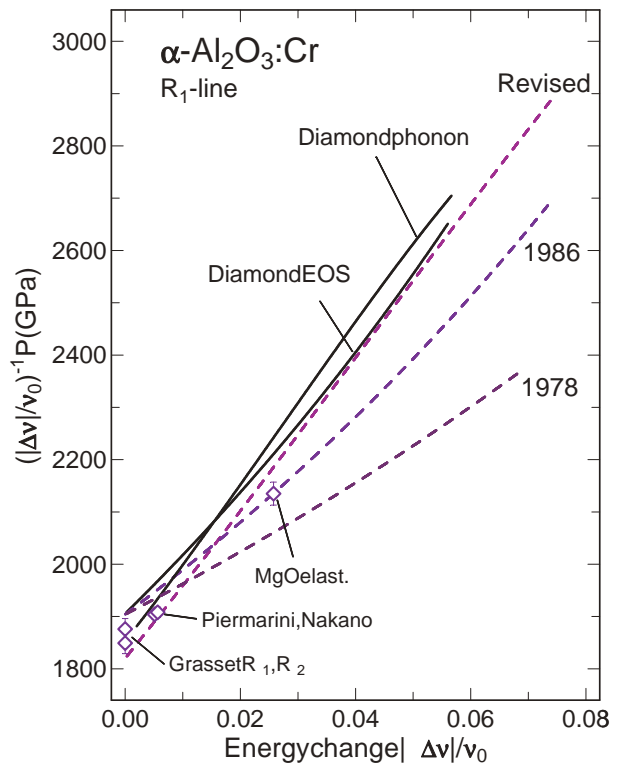}}
\caption[]{\label{fig6} Ruby pressure calibrations in reduced
coordinates. Pressure is divided by energy change of the R$_1$ line
and plotted as a function of the energy change. Dashed lines refer
to the calibrations by Mao \etal\cite{Mao78,Mao86} (marked 1976 and
1986) and the revision proposed by Holzapfel\cite{Holza02}. Lines
marked 'diamond' refer to converted data of Occelli
\etal\cite{Occelli02} combined with the calculated EOS and phonon
frequency shift of diamond reported in the present work. The
symbols near zero energy shift correspond to R1 line pressure
coefficients determined at low
pressures\cite{Pierm75,Nakan00,Grass01}. The symbol marked 'MgO'
stands for the result of elasticity studies of MgO up to 55
GPa\cite{Zha00}.}
\end{figure}

Holzapfel\cite{Holza02} recently proposed a revised ruby pressure
scale, based on an analysis of published EOS data of selected
elemental solids, including those of Ref.~[\onlinecite{Occelli02}]
for diamond. The revised ruby calibration was cast into a
three-parameter analytical expression for pressure as a function of
the R1 line wavelength. With the recommended parameter
values\cite{Holza02} and when restricted to about 200~GPa pressure
(a range beyond the upper limit of most DAC studies), the revised
calibration can be represented by a simple second-order polynomial
in frequency shift $\Delta \nu$ of the R1 line. This is evident
from Fig.~\ref{fig6} which shows plots of $(\vert \Delta \nu
\vert/\nu_0)^{-1} \times P$ versus $\vert \Delta \nu \vert /\nu_0$
for Holzapfel's revised scale and for the 1978/1986 scales of Mao
\etal\cite{Mao78,Mao86}~ We write the quadratic polynomial
as\cite{lambda}
\begin{equation}\label{eqsyas1}
P(\Delta \nu) = A\, \frac{\vert \Delta \nu\vert}{\nu_0}\left[1 +
\frac{B}{2} \frac{\vert \Delta \nu \vert}{\nu_0}\right]~.
\end{equation}
A linear regression for the 'Revised' line in Fig.~\ref{fig6} with
$A=1820$~GPa (fixed value as recommended in
Ref.~[\onlinecite{Holza02}]) gives a slope parameter of $B$=15.8.

From the experimental $P(V)$ and $P(\omega)$ data of
diamond\cite{Occelli02} one can recover the ruby line shift $\Delta
\nu$ according to the ruby calibration\cite{Mao86} used in the
experimental work. Combining the experimental $V(\Delta \nu)$ and
$\omega(\Delta \nu)$, respectively, with the calculated $P(V)$ and
$P(\omega)$ relations reported here yields $P(\Delta \nu)$ as shown
by the two solid lines in Fig.~\ref{fig6}. These curves run more or
less parallel to Holzapfel's revised calibration. Obviously,
\textit{both} the calculated $P(V)$ \textit{and} $P(\omega)$
relations for diamond in combination with the experimental data of
Ref.~[\onlinecite{Occelli02}] support the proposed revision of the
ruby pressure scale. It should be noted that the ruby calibration
discussed by Aleksandrov \etal\cite{Aleks87} would be a little
higher in pressure, but in the coordinates of Fig.~\ref{fig6} it
exhibits a slope similar to that of the 'revised' line.

\section{Conclusions}

We offer the following conclusions:

(1) The results of the \textit{first-principles} EOS and phonon
frequency calculations for diamond reported here do not depend on the
computational method, i.e., the choice of the crystal potential and
basis sets in the PW, PAW, and APW+lo methods.

(2) The calculated equilibrium values for volume ($V_0$) and bulk
modulus ($B_0$) of diamond do depend on the exchange-correlation
functional (LDA or GGA), a well-known fact in density functional
theory. However, the value of the pressure derivative of the bulk
modulus [$B_0^{\prime}$=3.65(5)] and the analytical form of the $P(V)$
relation are not affected within the volume range (40\% compression)
covered here. In other words, the calculated nonlinear component in the
$P(V)$ behavior is independent of the method of calculation and the
exchange-correlation approximation.

(3) The different theoretical methods and approximations also yield
very similar results for the nonlinearity in the pressure versus phonon
frequency relation $P(\omega)$.

(4) Both the calculated $P(V)$ \textit{and} $P(\omega)$ relations
exhibit a nonlinear behavior which is more pronounced than that
observed in recent x-ray diffraction and Raman measurements of diamond
up to 140~GPa\cite{Occelli02}. In terms of absolute pressure, the
differences between theoretical and experimental results are
essentially the same for the $P(V)$ and $P(\omega)$ behavior, of the
order of 10~GPa at an experimental pressure of 140~GPa.

(5) Our calculated nonlinearity of $P(\omega)$ agrees well with that
obtained in earlier \textit{ab initio} phonon frequency calculations
for diamond under pressure\cite{Hanfl85b,Niels86}. The work of Hanfland
\etal\cite{Hanfl85b} indicated the need to reconsider the calibration
of the ruby pressure scale used at that time\cite{Mao78}. The
'quasi-hydrostatic' (1986) ruby calibration\cite{Mao86} reduced the
discrepancy between theoretical predictions and experimental results.
The remaining difference discussed here at least calls for some caution
when experimental data measured above 50 GPa and based on the 1986 ruby
pressure scale are compared to \textit{ab initio} calculations or
reduced shock wave data.

\begin{acknowledgments}
The authors thank P. Loubeyre and W. B. Holzapfel for making
available preprints of Refs.~[\onlinecite{Occelli02}] and
[\onlinecite{Holza02}], respectively. Part of the computer
resources used in this work were provided by the Scientific
Committee of IDRIS (Institut du D\'{e}veloppement et des Ressources en
Informatique Scientifique), Orsay (France).
\end{acknowledgments}


\begin{thebibliography}{10}
\bibitem{lambda}
\bibinfo{note}{For practical purposes one may prefer to rewrite the $P(\nu)$
  relation in terms of wavelength: $$ P(\lambda) = A\, \frac{\Delta
  \lambda}{\lambda} \left[1 + \frac{B}{2} \frac{\Delta \lambda}{\lambda}
  \right]~.$$ Note that it is the wavelength at a given pressure which appears
  in the denominators.}
\end{thebibliography}

\printtables \printfigures

\end{document}